\documentclass[12pt,eqno,epsf]{article}
\usepackage{amsmath,amssymb,graphicx,slashbox}
\numberwithin{equation}{section}

\def\ignore#1{{}}

\newcounter{sxn}

\newcounter{axn}

\date{}

\newdimen\mybaselineskip
\renewcommand{\thefootnote}{\arabic{footnote}}

\newcommand{\beeq}{\begin{equation}}
\newcommand{\eneq}{\end{equation}}
\newcommand{\beqn}{\begin{eqnarray}}
\newcommand{\eeqn}{\end{eqnarray}}

\newcommand{\alp}{\alpha}
\newcommand{\bt}{\beta}

\newcommand{\dlt}{\delta}

\newcommand{\ep}{\epsilon}
\newcommand{\tht}{\theta}

\newcommand{\lmd}{\lambda}

\newcommand{\sgm}{\sigma}
\newcommand{\Sgm}{\Sigma}

\newcommand{\omg}{\omega}
\newcommand{\Omg}{\Omega}

\newcommand{\be}{\begin{equation}}
\newcommand{\ee}{\end{equation}}
\newcommand{\bea}{\begin{eqnarray}}
\newcommand{\eea}{\end{eqnarray}}
\newcommand{\eql}{\!\!\!&=\!\!\!&}

\newcommand{\sma}{\!\!\!&\simeq\!\!\!&}
\newcommand{\defa}{\!\!\!&\equiv\!\!\!&}

\newcommand{\simgt}{\stackrel{>}{{}_\sim}}
\newcommand{\simlt}{\stackrel{<}{{}_\sim}}

\newcommand{\tl}[1]{\tilde{#1}}
\newcommand{\bdm}[1]{{\mbox{\boldmath $#1$}}}

\newcommand{\tr}{{\rm tr}}

\newcommand{\der}{\partial}
\newcommand{\dr}{\!\!d}
\newcommand{\hc}{{\rm h.c.}}
\newcommand{\ie}{{i.e.}}
\newcommand{\id}{\mbox{\boldmath $1$}}

\newcommand{\vev}[1]{\langle #1 \rangle}
\newcommand{\Lvev}[1]{\left\langle #1 \right\rangle}
\newcommand{\brkt}[1]{\left( #1 \right)}
\newcommand{\brc}[1]{\left\{ #1 \right\}}
\newcommand{\sbk}[1]{\left[ #1 \right]}
\newcommand{\abs}[1]{\left| #1 \right|}

\renewcommand{\Re}{{\rm Re}\,}

\newcommand{\cA}{{\cal A}}

\newcommand{\cG}{{\cal G}}
\newcommand{\cH}{{\cal H}}

\newcommand{\cL}{{\cal L}}

\newcommand{\cN}{{\cal N}}
\newcommand{\cO}{{\cal O}}
\newcommand{\cP}{{\cal P}}
\newcommand{\cQ}{{\cal Q}}

\newcommand{\cV}{{\cal V}}

\newcommand{\cW}{{\cal W}}
\newcommand{\cX}{{\cal X}}
\newcommand{\cY}{{\cal Y}}
\newcommand{\cZ}{{\cal Z}}

\newcommand{\Io}{I_{\rm o}}
\newcommand{\Ie}{I_{\rm e}}
\newcommand{\Jo}{J_{\rm o}}
\newcommand{\Je}{J_{\rm e}}

\newcommand{\Ke}{K_{\rm e}}

\begin{document}
\thispagestyle{empty}

\baselineskip=12pt


\begin{flushright}
KEK-TH-1649 \\
WU-HEP-13-03 
\end{flushright}

\baselineskip=35pt plus 1pt minus 1pt

\vskip 1.5cm

\begin{center}
{\LARGE\bf Impacts of non-geometric moduli \\ on effective theory 
of 5D supergravity}

\vspace{1.5cm}
\baselineskip=20pt plus 1pt minus 1pt

\normalsize

{\large\bf Yutaka Sakamura}${}^{1,2}\!${\def\thefootnote{\fnsymbol{footnote}}
\footnote[1]{E-mail address: sakamura@post.kek.jp}} 
{\large\bf and Yusuke Yamada}${}^3\!${\def\thefootnote{\fnsymbol{footnote}}
\footnote[2]{E-mail address: yuusuke-yamada@asagi.waseda.jp}}

\vskip 1.0em

${}^1${\small\it KEK Theory Center, Institute of Particle and Nuclear Studies, 
KEK, \\ Tsukuba, Ibaraki 305-0801, Japan} \\ \vspace{1mm}
${}^2${\small\it Department of Particles and Nuclear Physics, \\
The Graduate University for Advanced Studies (Sokendai), \\
Tsukuba, Ibaraki 305-0801, Japan} 

\vskip 1.0em

${}^3${\small\it Department of Physics, Waseda University, \\ 
Tokyo 169-8555, Japan}

\end{center}

\vskip 1.0cm
\baselineskip=20pt plus 1pt minus 1pt

\begin{abstract}
5D supergravity generically has moduli other than 
the radion that belong to 5D vector multiplets. 
We summarize the impacts of such non-geometric moduli 
on 4D effective theory of 5D supergravity on $S^1/Z_2$. 
We mainly discuss the structure of the effective K\"ahler potential 
including the one-loop quantum corrections. 
As an illustrative example, we construct a model in which 
the size of the extra dimension is stabilized at an exponentially large value 
compared to the Planck length, which is similar to 
the LARGE volume scenario in string theory. 
\end{abstract}

\newpage

\section{Introduction}
Five-dimensional supergravity (5D SUGRA) compactified on an orbifold~$S^1/Z_2$ 
has been thoroughly investigated since it is the simplest setup for 
supersymmetric (SUSY) extra-dimensional models, and it can appear 
as an effective theory of string theory~\cite{Horava,Lukas}. 
Besides, SUSY extensions of the Randall-Sundrum model~\cite{Randall:1999ee} 
are also constructed in 5D SUGRA 
on $S^1/Z_2$~\cite{Gherghetta:2000qt,Falkowski:2000er,Altendorfer:2000rr}. 

When we construct phenomenological models based on 5D SUGRA, 
the size of the extra dimension must be stabilized at finite values. 
The corresponding modulus field is often referred to as the radion, 
and it belongs to a chiral multiplet in four-dimensional (4D) effective theories. 
Since the radion multiplet~$T_{\rm rad}$ often plays a significant role 
in the mediation of SUSY breaking to our visible sector, 
it is crucial to specify the $T_{\rm rad}$-dependence 
of the effective action. 
This issue was discussed in Refs.~\cite{Luty:1999cz,Luty:2000ec,Bagger:2000eh}. 

Here we should note that 5D SUGRA generically has other moduli 
whose masslessness at tree level is ensured by shift symmetries. 
Such moduli might also be the geometric moduli of 
more fundamental theories in higher dimensions, 
but from the viewpoint of 5D SUGRA, 
they are just massless scalar fields (at least at tree level). 
If such non-geometric moduli exist, they generically mix 
with the radion to form supermultiplets. 
The mixing is characterized by a cubic polynomial, which is referred to 
as the norm function~\cite{Kugo:2000af,Kugo:2002js,Kugo:2002vc}. 
This corresponds to the prepotential in 4D $N=2$ SUSY gauge theories. 
However, most models based on 5D SUGRA studied so far 
implicitly assumed that the non-geometric moduli do not exist. 
In our previous works~\cite{Abe:2004ar}-\cite{Sakamura:2013wia}, 
we have investigated various properties of 4D effective theories 
for {\it generic} 5D SUGRA, including the non-geometric moduli. 
However our discussions involve somewhat complicated technical aspects 
since we have worked in the superconformal 
formulation~\cite{Kugo:2000af,Kugo:2002js,Kugo:2002vc,Zucker} 
to deal with generic 5D SUGRA. 

In this paper, we would like to emphasize the impacts of the non-geometric moduli 
on effective theories of 5D SUGRA by 
summarizing the results obtained so far without going into technical details. 
Especially we focus on the differences from the conventional 
results~\cite{Luty:1999cz,Luty:2000ec,Bagger:2000eh} 
that do not include the non-geometric moduli. 
We also construct a simple model with multi moduli 
that dynamically realizes 
a large extra dimension~\cite{ArkaniHamed:1998rs} 
as an illustrative example. 
It is interesting to note that this model is similar to the LARGE volume scenario 
in string theory~\cite{Balasubramanian:2005zx,Conlon:2005ki} 
although no stringy effects are necessary in our model. 

The paper is organized as follows. 
In Sec.~\ref{setup}, we provide a compact review of 5D SUGRA action 
with boundary-localized terms. 
In Sec.~\ref{TreeLevel}, we show the results 
about the tree-level K\"ahler potential. 
Many of them have been obtained in our previous works, 
but it is worthwhile to summarize them in an organized way. 
In Sec.~\ref{Kahler:1loop}, 
we discuss the one-loop K\"ahler potential. 
We see that our formula is consistent with other related works, 
and construct a simple model that realizes the large extra dimension 
dynamically. 
Sec.~\ref{summary} is devoted to the summary. 
In Appendix~\ref{matrices}, we provide the definitions of 
matrices that characterize the vector sector.  
In Appendix~\ref{def:functions}, we listed the definitions of 
functions in the formula for the one-loop K\"ahler potential.

\section{Set-up} \label{setup}
We consider 5D SUGRA with generic prepotential, 
compactified on an orbifold~$S^1/Z_2$. 
We take the fundamental region of $S^1/Z_2$ as $0\leq y\leq L$, 
where $y$ is the coordinate of the extra dimension. 
The constant~$L$ is not the physical size of the extra dimension 
unless $\vev{e_y^{\;\;4}}=1$. 

\subsection{Field content}
5D SUGRA has the following supermultiplets,\footnote{
We do not consider the tensor multiplets, which are discussed 
in Refs.~\cite{Kugo:2002vc,Gunaydin:1999zx}. }
which are decomposed into $N=1$ superfields. 

\begin{description}
\item[Hypermultiplet] \mbox{}\\
A hypermultiplet is decomposed into two chiral superfields, 
which have opposite orbifold $Z_2$-parities. 
The hypermultiplets are divided into two classes, \ie,  
the compensator multiplet~$(\Phi_C,\Phi^c_C)$ 
and the physical matter multiplets~$(\cQ_a,\cQ_a^c)$, 
where the index~$a$ labels gauge multiplets. 
The former is an auxiliary multiplet and eliminated by 
the superconformal gauge fixing.\footnote{ 
In this paper, we assume that the number of the compensator multiplets is one, and have defined each $N=1$ superfield 
so that the Weyl weight of $\Phi_C$ is one 
and the others have zero weights. 
(See (A.2) in Ref.~\cite{Abe:2011rg}, for example.)
} 
The $Z_2$-parities of the $N=1$ superfields are shown in Table~\ref{Z2_parity}.\footnote{
We assume that each $N=1$ superfield has the same $Z_2$-parity 
at both boundaries, for simplicity. 
}  
Only the $Z_2$-even superfields~$(\Phi_C,\cQ_a)$ have 
zero-modes~$(\phi_C,Q_a)$.

\item[Vector multiplet] \mbox{}\\
A vector multiplet~$\mathbb V^I$ $(I=1,2,\cdots,n_V)$
is decomposed into $N=1$ vector and chiral superfields~$(V^I,\Sgm^I)$, 
which have opposite $Z_2$-parities. 
The vector multiplets are also divided into two classes 
according to their $Z_2$-parities. 
One is a class of the gauge supermultiplets, 
which are denoted as ${\mathbb V}^{\Ie}$ ($\Ie=1,\cdots,n_{V_{\rm e}}$). 
In this class, $V^{\Ie}$ are $Z_2$-even and have zero-modes 
that are identified with the gauge superfields in 4D effective theory. 
The other is a class of the moduli multiplets,  
which are denoted as ${\mathbb V}^{\Io}$ ($\Io=1,\cdots,n_{V_{\rm o}}$). 
In this class, the chiral superfields~$\Sgm^{\Io}$ have zero-modes~$T^{\Io}$, 
which are referred to as the moduli superfields in this paper. 
At least one vector multiplet belongs to the latter class,  
whose vector component is identified with the graviphoton. 

\end{description}
\begin{table}[t]
\begin{center}
\begin{tabular}{|c||c|c|c|c|c|c|c|c|c|c|} \hline
\rule[-2mm]{0mm}{7mm} 5D multiplet & \multicolumn{4}{c|}{Hypermultiplet} 
& \multicolumn{4}{c|}{Vector multiplet} \\ \hline
Role & \multicolumn{2}{c|}{compensator} 
& \multicolumn{2}{c|}{matter} 
& \multicolumn{2}{c|}{moduli} & \multicolumn{2}{c|}{gauge} \\ \hline
$N=1$ superfield & \mbox{ } $\Phi_C$ \mbox{ } & $\Phi_C^c$ & $\cQ_a$ & $\cQ_a^c$ & 
$V^{\Io}$ & $\Sgm^{\Io}$ & $V^{\Ie}$ & $\Sgm^{\Ie}$ \\ \hline
$Z_2$-parity & $+$ & $-$ & $+$ & $-$ & $-$ & $+$ & $+$ & $-$ \\ \hline
Zero-mode & $\phi_C$ & & $Q_a$ & & & $T^{\Io}$ & $V^{\Ie}$ & \\ \hline
\end{tabular}
\end{center}
\caption{The decomposition of 5D supermultiplets into $N=1$ superfields. 
The orbifold $Z_2$-parities of the $N=1$ superfields are also shown. 
Only the $Z_2$-even superfields have zero-modes that will appear 
in 4D effective theory. 
}
\label{Z2_parity}
\end{table}

Besides the above supermultiplets, we have the gravitational multiplet, 
or the Weyl multiplet in the superconformal formulation. 
It has to be taken into account when the one-loop correction 
is evaluated~\cite{Sakamura:2013wia}. 

The gauge-invariant field-strength superfields are defined as~\footnote{
Note that $\cV$ is not hermitian, but $e^{-\frac{V}{2}}\cV e^{\frac{V}{2}}$ is. 
} 
\bea
 \cW_\alp \defa \frac{1}{4}\bar{D}^2\brkt{e^V D_\alp e^{-V}}+\cdots, \nonumber\\
 \cV \defa e^V\der_y e^{-V}+\Sgm+e^V\Sgm^\dagger e^{-V}+\cdots, 
\eea
where the ellipses denote terms involving the gravitational 
superfields~\cite{Sakamura:2012bj}. 
We have used a matrix notation~$(V,\Sgm)\equiv (V^I,\Sgm^I)t_I$, 
where the hermitian generators~$t_I$ act on $(\Phi_C,\cQ_a)$ 
and contain gauge coupling constants. 
The gauge supermultiplets~$(V_{\rm e},\Sgm_{\rm e})\equiv (V^{\Ie},\Sgm^{\Ie})t_{\Ie}$ 
are divided into the sum of the matrices~$(V_r,\Sgm_r)$, 
where $r$ labels the simple or Abelian factors of the gauge group. 
In this paper, we assume that the gauge groups 
for the moduli multiplets~${\mathbb V}^{\Io}$ are Abelian, 
and $\Phi_C$ and $\cQ_a$ have charges~$-3k_{\Io}$ and $-2d_{a\Io}$ for ${\mathbb V}^{\Io}$, 
respectively. 
Namely, the corresponding generators~$t_{\Io}$ are 
\be
 t_{\Io} = -3k_{\Io}\cP_C \oplus\bigoplus_a\brkt{-2d_{a\Io}\otimes\id_{n_a}}. 
\ee
where $\cP_C$ is a projection operator onto $\Phi_C$, and 
$n_a$ denotes the dimension of the gauge-group representation 
that $\cQ_a$ belongs to. 
The gauge coupling constants~$k_{\Io}$ and $d_{a\Io}$ induce 
the 5D cosmological constant and 5D bulk masses for $(\cQ_a,\cQ_a^c)$, 
respectively.
These coupling constants are $Z_2$-odd. 
Such kink-type couplings can be realized in SUGRA context 
by the mechanism proposed in Ref.~\cite{Bergshoeff:2000zn}. 

The vector sector is characterized by a cubic polynomial~$\cN(\cV)$, 
which is referred to as the norm function. 
This is defined as 
\be
 \cN(\cV) \equiv \frac{c_{\rm vc}}{3}\tr\brkt{t_I\brc{t_J,t_K}}\cV^I\cV^J\cV^K, 
 \label{def:norm}
\ee
where the real constant~$c_{\rm vc}$ can take different values 
for each simple or Abelian factor of the gauge group. 
Eq.(\ref{def:norm}) must be $Z_2$-even,  
and thus can be rewritten in the following form. 
\be
 \cN(\cV) = \sum_r C^r_{\Io}\cV^{\Io}\tr(\cV_r^2)
 +\hat{\cN}(\cV^{\Io}), \label{def:norm_fcn}
\ee
where $C_{\Io}^r$ are real constants, and 
the second term is a cubic function of only $\cV^{\Io}$.

\subsection{5D Lagrangian}
Since only $N=1$ SUSY is preserved by the orbifold projection, 
it is convenient to express 5D Lagrangian 
in terms of the $N=1$ superfields in Table~\ref{Z2_parity}. 
It is an extension of Ref.~\cite{ArkaniHamed:2001tb} to the local SUSY case. 
Besides couplings to the gravitational superfields,\footnote{
Terms involving the gravitational superfields are listed in Ref.~\cite{Sakamura:2012bj}. 
} the 5D Lagrangian is expressed as~\cite{Abe:2006eg,Correia:2006pj}
\bea
 \cL \eql -\int\dr^4\tht\;3\abs{\Phi_C}^2\cN^{1/3}(\cV)e^{2k\cdot\tl{V}} \nonumber\\
 &&\hspace{10mm}\times
 \left\{1+e^{-6k\cdot V}\abs{\Phi_C^c}^2 
 -\sum_a\brkt{e^{2d_a\cdot V}\cQ_a^\dagger e^{-V_{\rm e}}\cQ_a
 +e^{-2d_a\cdot V}\cQ_a^{c\dagger}(e^{V_{\rm e}})^t\cQ_a^c}\right\}^{2/3} \nonumber\\
 &&+\sbk{\int\dr^2\tht\;2\Phi_C^3\brc{\brkt{\der_y-3k\cdot\Sgm}\Phi_C^c
 -\sum_a\cQ_a^t\brkt{\der_y-(2d_a+3k)\cdot\Sgm+\Sgm_{\rm e}^t}\cQ_a^c}+\hc} \nonumber\\
 &&+\cL_{\rm vc}+2\sum_{y_*=0,L}\cL_{\rm bd}^{(y_*)}\dlt(y-y_*),  \label{5D_cL}
\eea
where 
$k\cdot V\equiv k_{\Io}V^{\Io}$, $d_a\cdot V\equiv d_{a\Io}V^{\Io}$, $\cdots$, and  
$\cL_{\rm vc}$ contains the kinetic term for $V$ and the Chern-Simons term. 
In the Wess-Zumino gauge, $\cL_{\rm vc}$ is written 
as~\cite{ArkaniHamed:2001tb,Hebecker:2008rk} 
\bea
 \cL_{\rm vc} \defa -\int\dr^2\tht\;c_{\rm vc}
 \tr\sbk{\Sgm\cW^2-\frac{1}{24}\bar{D}^2\brkt{
 \brc{V,\der_y D^\alp V}-\brc{\der_y V,D^\alp V}}
 \brkt{\cW_\alp-\frac{1}{4}\cW_\alp^{(2)}}}  \nonumber\\
 &&+\hc, 
\eea 
where $\cW_\alp^{(2)}$ is a quadratic part of $\cW_\alp$ in $V$.  
The fractional powers in (\ref{5D_cL}) appear after integrating out 
an auxiliary superfield in the 5D gravitational multiplet~\cite{Correia:2006pj}. 

The boundary Lagrangian~$\cL_{\rm bd}^{(y_*)}$ ($y_*=0,L$) is expressed as 
\bea
 \cL_{\rm bd}^{(y_*)} \eql -\sbk{\int\dr^2\tht\;
 \sum_r\frac{1}{2}f^{(y_*)r}(\cQ,q_{y_*})\tr\brkt{\cW_r^2}+\hc} \nonumber\\
 &&+\int\dr^4\tht\;\abs{\Phi_C}^2 
 \Omg^{(y_*)}(\cQ,q_{y_*},V)
 +\sbk{\int\dr^2\tht\;\Phi_C^3 W^{(y_*)}(\cQ,q_{y_*})+\hc}, \label{cL_bd}
\eea
where $f^{(y_*)r}$ and $W^{(y_*)}$ are holomorphic functions 
and $\Omg^{(y_*)}$ is a real function. 
The bulk superfields are evaluated at $y=y_*$, and 
$q_{y_*}$ denotes 4D chiral superfields localized at $y=y_*$. 

The warp factor does not appear explicitly in the above expressions 
since it can be absorbed by using the dilatation, which is 
a part of the superconformal symmetry that is respected in our formulation.

\section{Effective theory at tree level} \label{TreeLevel}
Following the procedure developed in Refs.~\cite{Abe:2006eg,Abe:2008an,Abe:2011rg}, 
we can derive 4D effective Lagrangian at tree level for the 5D theory~(\ref{5D_cL}), 
which is expressed in the following form. 
\bea
 \cL_{\rm eff} \eql -\sbk{\int\dr^2\tht\;
 \sum_r \frac{1}{2}f_{\rm eff}^r(Q,T)\tr\brkt{\cW_r^2}+\hc} \nonumber\\
 &&+\int\dr^4\tht\;\abs{\phi_C}^2\Omg_{\rm eff}(|\tl{Q}|^2,\Re T)
 +\sbk{\int\dr^2\tht\;\phi_C^3 W_{\rm eff}(Q,T)+\hc},  \label{cL_eff}
\eea
where $|\tl{Q}_a|^2\equiv Q_a^\dagger e^{-V_{\rm e}}Q_a$. 
Here we omit the 4D boundary superfields~$q_{y_*}$ in this section. 
The effective gauge kinetic functions and superpotential are given by 
\bea
 f_{\rm eff}^r(Q,T) \eql C_{\Io}^r T^{\Io}+f^{(0)r}(Q)+f^{(L)r}(e^{-d_a\cdot T}Q_a), 
 \nonumber\\
 W_{\rm eff}(Q,T) \eql W^{(0)}(Q)+e^{-3k\cdot T}W^{(L)}(e^{-d_a\cdot T}Q_a).  
 \label{eff:fW}
\eea
The matter fields~$Q_a$ can appear in $f^{(y_*)r}$ 
only through the gauge-singlet combinations. 
The effective K\"ahler potential~$\Omg_{\rm eff}=-3e^{-K_{\rm eff}/3}$ 
has a more complicated structure. 
We will explain it in the next two subsections. 

Here let us comment on the physical size of the extra dimension~$L_{\rm phys}$. 
The relation between the 4D and 5D Planck masses~$M_{\rm Pl}$ and $M_5$ 
is $M_{\rm Pl}^2=M_5^3L_{\rm phys}$.\footnote{
We neglect the contributions from the boundary terms.} 
In the 4D Einstein frame, the compensator scalar~$\phi_{C}$ is fixed as 
$\vev{\abs{\phi_{C}}^2\Omg_{\rm eff}}=-3M_{\rm Pl}^2$~\cite{Sakamura:2011df}. 
On the other hand, the gauge-fixing condition for the 5D Einstein frame 
is $\vev{\abs{\Phi_{C}}^2}=M_5^3$~\cite{Abe:2004ar}. 
In our derivation of (\ref{cL_eff}), it follows that 
$\vev{\phi_{C}}=\vev{\Phi_{C}^{2/3}}$~\cite{Abe:2006eg}. 
Therefore, we find that $L_{\rm phys}=(-\vev{\Omg_{\rm eff}}/3)^{3/2}/M_{\rm Pl}$. 
In the rest of this paper, we will basically take a unit of $M_{\rm Pl}$.

\subsection{Flat spacetime}
The flat 5D spacetime is realized when the compensator multiplet is neutral 
for the moduli multiplets, \ie, $k_{\Io}=0$. 

In the single modulus case~($n_{V_{\rm o}}=1$), 
the effective K\"ahler potential~$\Omg_{\rm eff}$ can be easily 
calculated as~\cite{Correia:2006pj} 
\bea
 \Omg_{\rm eff} \eql \Omg^{(0)}(Q,V)
 +\Omg^{(L)}(e^{-d_a T_{\rm rad}}Q_a,V) \nonumber\\
 &&-3\Re T_{\rm rad}
 +\sum_a \frac{1-e^{-2d_a\Re T_{\rm rad}}}{d_a}|\tl{Q}_a|^2
 +\sum_{a,b}\frac{1-e^{-2(d_a+d_b)\Re T_{\rm rad}}}
 {6(d_a+d_b)}|\tl{Q}_a|^2|\tl{Q}_b|^2 \nonumber\\
 &&+\cO(Q^6),  \label{single:flat}
\eea
where $d_a\equiv d_{a1}$ are the bulk masses for $(\cQ_a,\cQ^c_a)$ in the unit of 
$M_5$, 
and $T_{\rm rad}\equiv T^1$ is the radion superfield. 
The first line is the contributions of the boundary terms, and 
the second line is from the bulk. 
The first term in the second line is the well-known radion K\"ahler 
potential~\cite{Luty:1999cz}. 
The second term is relevant to the realization of the hierarchy among 
Yukawa couplings for the matter fields~$Q_a$ 
after the canonical normalization. 
The third term contributes to the soft SUSY-breaking masses 
when some of $Q_b$ are SUSY-breaking superfields with  
non-vanishing F-terms.\footnote{
The induced soft masses turn out to be tachyonic 
in the single modulus case~\cite{Abe:2008ka}. 
Related issues are also discussed in Ref.~\cite{Dudas:2012mv}. 
} 

In the multi moduli case~($n_{V_{\rm o}}\geq 2$), 
(\ref{single:flat}) is modified as 
\bea
 \Omg_{\rm eff} \eql \Omg^{(0)}(Q,V)+\Omg^{(L)}(e^{-d_a\cdot T}Q_a,V) \nonumber\\
 &&-3\hat{\cN}^{1/3}\left\{
 1-\frac{2}{3}\sum_a Y_{d_a}|\tl{Q}_a|^2 
 +\sum_{a,b}\tl{\Omg}_{a,b}^{(4)}|\tl{Q}_a|^2|\tl{Q}_b|^2\right\}+\cO(Q^6), 
 \label{multi:flat}
\eea
where 
\bea
 Y_d(\Re T) \defa \frac{1-e^{-2d\cdot\Re T}}{2d\cdot\Re T}, \nonumber\\
 \tl{\Omg}_{a,b}^{(4)}(\Re T) \defa 
 \frac{(d_a\cdot\cP_V a^{-1}\cdot d_b)\brc{Y_{d_a+d_b}-Y_{d_a}Y_{d_b}}}
 {3(d_a\cdot\Re T)(d_b\cdot\Re T)}-\frac{Y_{d_a+d_b}}{9}. 
\eea
A matrix~$a_{IJ}$ and a projection operator~$\cP_V$ 
are defined in (\ref{def:a_IJ}) and (\ref{def:cP_V}), respectively~\cite{Kugo:2000af}. 
Here and henceforth, the arguments of the norm function~$\hat{\cN}$ and its derivatives 
are understood as $(\Re T^{\Io})$. 
The radion superfield~$T_{\rm rad}$ is identified as 
\be
 T_{\rm rad} = \Lvev{\frac{\hat{\cN}_{\Io}}{3\hat{\cN}^{2/3}}}T^{\Io}, 
 \label{def:T_rad}
\ee
where $\vev{\cdots}$ denotes the vacuum expectation value (VEV). 
The effective theory in this case has the following properties. 
\begin{itemize}
\item 
Since the size of the extra dimension is 
$L_{\rm phys}=\vev{\hat{\cN}^{1/2}}=(\Re\vev{T_{\rm rad}})^{3/2}$, 
the Kaluza-Klein (KK) mass scale 
is provided by $m_{\rm KK}=\pi/\vev{\hat{\cN}^{1/2}}$, 
which is regarded as the cutoff scale of 4D effective theory. 

\item
Note that the matter-independent part~$\Omg_{\rm eff}^{\rm moduli}=-3\hat{\cN}^{1/3}$ 
is expanded as 
\bea
 \Omg_{\rm eff}^{\rm moduli} \eql 
 -3\left\{\vev{\hat{\cN}^{1/3}}
 +\Lvev{\frac{\hat{\cN}_{\Io}}{3\hat{\cN}^{2/3}}}\Re\tl{T}^{\Io} 
 \right.\nonumber\\
 &&\left.\hspace{10mm}
 +\Lvev{\frac{3\hat{\cN}\hat{\cN}_{\Io\Jo}-2\hat{\cN}_{\Io}\hat{\cN}_{\Jo}}
 {18\hat{\cN}^{5/3}}}(\Re\tl{T}^{\Io})
 (\Re\tl{T}^{\Jo})+\cO\brkt{(\Re\tl{T})^3}\right\} \nonumber\\
 \eql -3\Re T_{\rm rad}+\Lvev{\hat{\cN}^{1/3}(a\cdot\cP_V)_{\Io\Jo}}
 (\Re T^{\Io})(\Re T^{\Jo})+\cO\brkt{(\Re\tl{T})^3}, \label{Omg^moduli}
\eea
where $\tl{T}^{\Io}\equiv T^{\Io}-\vev{T^{\Io}}$. 
The first term is the radion K\"ahler potential, which already appeared 
in the single modulus case~(\ref{single:flat}),  
while the second term represents the kinetic terms for the non-geometric moduli. 
As we can see from (\ref{Omg^moduli}), 
if we single out the radion~$T_{\rm rad}$ 
from the moduli~$T^{\Io}$ and treat it separately, 
the projection operator~$\cP_V$ appears in the kinetic terms 
for the other moduli, which makes awkward to treat them. 
Thus it is convenient to treat all the moduli on equal footing 
in the multi moduli case. 

\item
The moduli K\"ahler potential~$K^{\rm moduli}_{\rm eff}
=-3\ln(-\Omg_{\rm eff}^{\rm moduli}/3)=-\ln\hat{\cN}$, 
and has the no-scale structure. 
Thus the potential for the moduli is not generated at tree level. 
The one-loop correction breaks this structure, 
as we will see in the next section. 

\item
The first term in $\tl{\Omg}_{a,b}^{(4)}$ 
is peculiar to the multi moduli case. 
It is induced by integrating out the non-geometric moduli, 
and can significantly affect the sfermion masses for the bulk matter fields. 
In fact, we have pointed out that non-tachyonic and approximately flavor universal 
sfermion masses are naturally obtained when the fermion mass hierarchy is realized  
by the wave function localization~\cite{Abe:2008an}. 
\end{itemize}

\subsection{Warped spacetime}
Next we consider a case that the compensator multiplet is charged 
for the moduli multiplets, \ie, $k_{\Io}\neq 0$. 
In this case, the background spacetime has a nontrivial warped geometry. 

In the single modulus case, it becomes the Randall-Sundrum 
spacetime~\cite{Gherghetta:2000qt,Falkowski:2000er,Altendorfer:2000rr}, 
\be
 ds^2 = e^{-2ky}\eta_{\mu\nu}dx^\mu dx^\nu-dy^2, 
 \;\;\;\;\;(\mu,\nu=0,1,2,3)  \label{RSmetric}
\ee
where $k\equiv k_1$ is the AdS curvature scale in the unit of $M_5$. 
The effective K\"ahler potential~$\Omg_{\rm eff}$ is obtained as 
\bea
 \Omg_{\rm eff} \eql \Omg^{(0)}(Q,V)
 +e^{-2k\Re T_{\rm rad}}\Omg^{(L)}(e^{-d_a T_{\rm rad}}Q_a,V)  \nonumber\\
 &&-3\Re T_{\rm rad}\brc{Y_k-\sum_a \frac{2Y_{k+d_a}}{3}|\tl{Q}_a|^2
 -\sum_{a,b}\frac{Y_{k+d_a+d_b}}{9}|\tl{Q}_a|^2|\tl{Q}_b|^2}+\cO(Q^6). 
 \label{single:warp}
\eea
The matter-independent term in the second line is 
the well-known radion K\"ahler potential~\cite{Luty:2000ec,Bagger:2000eh}. 
The expression~(\ref{single:warp}) will reduce to (\ref{single:flat}) 
in the limit~$k\to 0$ since $\lim_{k\to 0}Y_k=1$. 

In the multi moduli case, on the other hand, 
it is much more difficult to derive $\Omg_{\rm eff}$. 
In our previous work~\cite{Abe:2011rg}, we derived it under the condition, 
\be
 \vev{k\cdot\cP_V} = 0, \label{kcP}
\ee
as 
\bea
 \Omg_{\rm eff} \eql \Omg^{(0)}(Q,V)
 +e^{-2k\cdot\Re T}\Omg^{(L)}(e^{-d_a\cdot T}Q_a,V) \nonumber\\
 &&-3\hat{\cN}^{1/3}\left\{Y_k
 -\sum_a \frac{2Y_{k+d_a}}{3}|\tl{Q}_a|^2
 +\sum_{a,b}\tl{\Omg}_{a,b}^{(4)}|\tl{Q}_a|^2|\tl{Q}_b|^2\right\}+\cO(Q^6), 
 \label{Omg^1loop:RS}
\eea
where
\be
 \tl{\Omg}_{a,b}^{(4)} = \frac{(d_a\cdot\cP_V a^{-1}\cdot d_b)
 \brc{Y_{k+d_a+d_b}-\frac{Y_{d_a}Y_{d_b}}{Y_{-k}}}}
 {3\brc{(k+d_a)\cdot\Re T}\brc{(k+d_b)\cdot\Re T}}
 -\frac{Y_{k+d_a+d_b}}{9}.  \label{multi:RS}
\ee
The corresponding background geometry is 
the Randall-Sundrum spacetime~(\ref{RSmetric}). 
The first term in (\ref{multi:RS}) is peculiar to the multi moduli case, 
just like in the case of the flat spacetime. 
The condition~(\ref{kcP}) indicates that the compensator is charged 
only for the graviphoton~(\ref{def:V_G}), 
which is the $N=2$ superpartner for the radion~(\ref{def:T_rad}). 
However, this condition does not seem to be natural 
because it is a relation between the parameters of the theory~$k_{\Io}$ 
and the VEVs of the moduli. 
This means that the Randall-Sundrum spacetime is only a special limit 
in the multi moduli case. 
This is because the background field configuration of the non-geometric moduli 
generically contribute to the spacetime geometry. 
We need some moduli-stabilization mechanism that realizes (\ref{kcP}) 
dynamically in order to justify (\ref{Omg^1loop:RS}). 

For arbitrary choices of $k_{\Io}$, 
an explicit form of $\Omg_{\rm eff}$ is known only 
in the case that the norm function is monomial. 
Thus consider a case that 
\be
 \hat{\cN}(\cV) = (\cV^1)^2\cV^2.  \label{mono_norm}
\ee
Then we obtain the moduli K\"ahler potential as~\cite{Correia:2006pj} 
\be
 \Omg_{\rm eff} = -3\hat{\cN}^{1/3}Y^{2/3}_{\frac{3}{2}k_1}Y^{1/3}_{3k_2}
 +\cO(Q^2), 
\ee
and the corresponding background geometry is 
\bea
 ds^2 \eql e^{2\sgm(y)}\eta_{\mu\nu}dx^\mu dx^\nu-e^{-4\sgm(y)}dy^2, \nonumber\\
 e^{2\sgm(y)} \eql \brkt{3k_1y+\Lvev{\frac{3k_1\hat{\cN}^{1/3}}
 {e^{3k_1\Re T^1}-1}}}^{2/3}
 \brkt{6k_2y+\Lvev{\frac{6k_2\hat{\cN}^{1/3}}{e^{6k_2\Re T^2}-1}}}^{1/3}. 
 \label{metric:2moduli}
\eea
Since $L_{\rm phys}=-\vev{\Omg_{\rm eff}}/3$, the KK mass scale is 
$m_{\rm KK}=\pi/\vev{\hat{\cN}^{1/3}Y^{2/3}_{\frac{3}{2}k_1}Y^{1/3}_{3k_2}}$. 
The condition~(\ref{kcP}) now becomes $k_1\Re\vev{T^1}=2k_2\Re\vev{T^2}$ 
(see (\ref{expr:acP_V})), and under this condition, 
(\ref{metric:2moduli}) becomes 
the Randall-Sundrum metric~(\ref{RSmetric}) after the coordinate redefinition. 

The matter-dependent terms are more complicated even in the case 
of the simple norm function~(\ref{mono_norm}), 
but they are reduced to simpler forms in some limits. 
In the $k_2=d_{a2}=0$ case, the bulk contribution to $\Omg_{\rm eff}$ is obtained as 
\bea
 \Omg_{\rm eff}^{\rm bulk} \eql 
 -3\hat{\cN}^{1/3}Y^{2/3}_{\frac{3}{2}k_1}  \label{multi:warp1} \\
 &&\times\left\{1-\sum_a\frac{2Y_{d_{a1}+\frac{3}{2}k_1}}{3Y_{\frac{3}{2}k_1}}|\tl{Q}_a|^2 
 -\sum_{a,b}\frac{Y_{d_{a1}+\frac{3}{2}k_1}Y_{d_{b1}
 +\frac{3}{2}k_1}}{9Y^2_{\frac{3}{2}k_1}}|\tl{Q}_a|^2|\tl{Q}_b|^2\right\}
 +\cO(Q^6),  \nonumber
\eea
and in the $k_1=d_{a1}=0$ case, it becomes 
\bea
 \Omg_{\rm eff}^{\rm bulk} \eql -3\hat{\cN}^{1/3}Y^{1/3}_{3k_2}
 \left\{1-\sum_a\frac{2Y_{d_{a2}+3k_2}}{3Y_{3k_2}}|\tl{Q}_a|^2 
 +\sum_{a,b}\tl{\Omg}_{a,b}^{(4)}|\tl{Q}_a|^2|\tl{Q}_b|^2\right\}+\cO(Q^6), 
 \label{multi:warp2} 
\eea
where 
\be
 \tl{\Omg}_{a,b}^{(4)} = \frac{Y_{d_{a2}+d_{b2}+3k_2}}{3Y_{3k_2}}
 -\frac{4Y_{d_{a2}+3k_2}Y_{d_{b2}+3k_2}}{9Y^2_{3k_2}}. 
\ee
In the flat limit~$k_1,k_2\to 0$, 
both (\ref{multi:warp1}) and (\ref{multi:warp2}) are reduced to 
(\ref{multi:flat}) with the norm function~(\ref{mono_norm}). 
For arbitrary gaugings, the matter-dependent part of $\Omg_{\rm eff}^{\rm bulk}$ 
becomes much more complicated, and is calculated as 
\bea
 \Omg_{\rm eff}^{\rm bulk} \eql -3\cN^{1/3}Y^{2/3}_{\frac{3}{2}k_1}Y^{1/3}_{3k_2}
 \sbk{1-\sum_a\frac{2\cX}{3Y_{\frac{3}{2}k_1}}|\tl{Q}_a|^2+\cO(Q^4)}, 
 \label{multi:warp3}
\eea
where 
\bea
 \cX(\Re T) \defa -\int_0^{-\Re T^1}\dr U\;\frac{e^{(2d_{a1}+3k_1)U}}{\Re T^1}
 \brkt{\frac{\cA+e^{3k_1U}}{\cA+1}}^{\frac{d_{a2}}{3k_2}} \nonumber\\
 \eql -\frac{1}{(2d_{a1}+3k_1)\Re T^1}
 \brkt{\frac{\cA}{\cA+1}}^{\frac{d_{a2}}{3k_2}} \nonumber\\
 &&\times\sbk{e^{(2d_{a1}+3k_1)U}
 {}_2F_1\brkt{1+\frac{2d_{a1}}{3k_1},-\frac{d_{a2}}{3k_2},
 2+\frac{2d_{a1}}{3k_1};-\frac{e^{3k_1U}}{\cA}}}_0^{-\Re T^1}, \nonumber\\
 \cA(\Re T) \defa \frac{e^{-6k_2\Re T^2}-e^{-3k_1\Re T^1}}{1-e^{-6k_2\Re T^2}}. 
\eea
Here ${}_2F_1(a,b,c;z)$ is the hypergeometric function. 
Notice that
\be
 \lim_{k_1,k_2\to 0}\cX = Y_{d_a}, \;\;\;\;\;
 \lim_{k_2,d_{a2}\to 0}\cX = Y_{d_{a1}+\frac{3}{2}k_1}, 
 \;\;\;\;\;
 \lim_{k_1,d_{a1}\to 0}\cX = \frac{Y_{d_{a2}+3k_2}}{Y_{3k_2}},  
\ee
and thus (\ref{multi:warp3}) is consistent 
with the results~(\ref{multi:flat}), (\ref{multi:warp1}) and (\ref{multi:warp2}). 

Besides the case of (\ref{mono_norm}), 
there are some other cases in which $\Omg_{\rm eff}$ can be calculated. 
For example, the norm function,  
\be
 \hat{\cN}(\cV) = \alp(\cV^1)^3+\bt(\cV^1)^2\cV^2, 
\ee
where $\alp$ and $\bt$ are arbitrary real constants, 
reduces to the form of (\ref{mono_norm}) by the field redefinition, 
\be
 \tl{\cV}^1 = \cV^1, \;\;\;\;\;
 \tl{\cV}^2 = \alp\cV^1+\bt\cV^2. 
\ee
Thus, $\Omg_{\rm eff}$ can be obtained by replacing $(T^1,T^2)$ 
with $(T^1,\alp T^1+\bt T^2)$ in the above expressions.

\section{One-loop K\"ahler potential} \label{Kahler:1loop}
In this section, we discuss the one-loop contribution to $\Omg_{\rm eff}$. 
We consider the case of flat spacetime, 
in which the tree-level K\"ahler potential has the no-scale structure. 
The one-loop correction breaks such structure, 
and generate the potential for the moduli, which is necessary for 
the moduli stabilization.  
Here we will focus on this property and 
neglect terms involving the bulk matter multipltets in $\Omg_{\rm eff}^{\rm 1loop}$. 
Such terms only provide subleading corrections to the counterparts 
in the tree-level K\"ahler potential.

\subsection{General expression}
We have derived the one-loop contribution to $\Omg_{\rm eff}$ 
for arbitrary forms of the norm function in Ref.~\cite{Sakamura:2013wia}. 
The result is~\footnote{
If we calculate $\Omg_{\rm eff}^{\rm 1loop}$ on an interval~$0\leq y\leq L$, 
we have divergent terms proportional to $(d_a\cdot\Re T)^3$. 
Such terms are canceled for theories on $S^1/Z_2$ 
with the contribution for $-L\leq y\leq 0$ 
because they are $Z_2$-odd. 
} 
\bea
 \Omg_{\rm eff}^{\rm 1loop} \eql \frac{1}{8\pi^2\hat{\cN}^{2/3}}\left[
 (n_V+1)\cZ(0)-\sum_a n_a\cZ\brkt{d_a\cdot\Re T}
 \right.\nonumber\\
 &&\hspace{20mm}\left.
 +\int_0^\infty\dr\lmd\sum_{F=U,V,{\rm ch}}g_F\lmd
 \ln\frac{\cG_F(\lmd)}{\cH_F^{(L)}(\lmd)\cH_F^{(0)}(\lmd)}\right]+\cO(Q^2), 
 \label{Omg^1loop}
\eea
where $n_a$ denotes the dimension of the gauge-group representation 
that $Q_a$ belongs to, 
$(g_U,g_V,g_{\rm ch})=(-2,-1,\frac{1}{2})$, 
the functions~$\cG_F$ and $\cH_F^{(y_*)}$ 
are listed in Appendix~\ref{def:functions}, and 
\bea
 \cZ(x) \defa -\int_0^\infty\dr\lmd\;\lmd
 \ln\brkt{2e^{-\sqrt{\lmd^2+x^2}}\sinh\sqrt{\lmd^2+x^2}}.  
\eea
The brane-to-brane loop effects are contained in 
the second line of (\ref{Omg^1loop}), and $F=U,V,{\rm ch}$ denote 
the contributions of the loops of the gravitational, vector and chiral multiplets, 
respectively. 
The function~$\cZ(x)$ is an even function that has a maximum value 
$\cZ(0)=\zeta(3)/4\simeq 0.30$ at $x=0$ 
and exponentially decreases as $\abs{x}$ increases. 
In fact, it is negligible when $\abs{x}\simgt 3$. 
In the case that $\abs{d_a\cdot\Re T}\gg 1$, the wave function for $Q_a$ 
strongly localized toward one of the boundaries. 
So only the zero-modes that spread over the bulk contribute to 
the first line in (\ref{Omg^1loop}).

The above expression is consistent with the results 
in Refs~\cite{Falkowski:2005fm,Rattazzi:2003rj,Gregoire:2004nn}. 
To see this, let us 
consider a simple case that there are no non-geometric moduli ($n_{V_{\rm o}}=1$), 
no bulk masses for the hypermultiplets ($d_{a}=0$), 
no gauge kinetic terms nor superpotentials at the boundaries ($f^{(y_*)}=W^{(y_*)}=0$), 
and $\Omg^{(y_*)}$ only depend on the localized chiral multiplets~$q_{y_*}$.   
Then (\ref{Omg^1loop}) becomes 
\be
 \Omg_{\rm eff}^{\rm 1loop} = \frac{1}{8\pi^2(\Re T_{\rm rad})^2}\left[
 (n_{V_{\rm e}}-n_H+2)\frac{\zeta(3)}{4}
 -\int_0^\infty\dr\lmd\;2\lmd\ln
 \frac{\cG_U(\lmd)}{\cH_U^{(L)}(\lmd)\cH_U^{(0)}(\lmd)}\right]+\cO(Q^2), 
\ee
where $n_H\equiv \sum_a n_a$ is the number of the physical hypermultiplets. 
The coefficient of the bulk contribution~$(n_{V_{\rm e}}-n_H+2)$ is consistent with 
(4.7) of Ref.~\cite{Falkowski:2005fm}. 
Now we further assume that there are no bulk matter fields, \ie, $n_{V_{\rm e}}=n_H=0$. 
Since 
\be
 \int_0^\infty\dr\lmd\;\lmd\ln\frac{\cG_U}{\cH_U^{(L)}\cH_U^{(0)}} 
 = \frac{\zeta(3)}{4}+\int_0^\infty\dr\lmd\;\lmd\ln\brc{1
 -\frac{(1-\frac{\lmd\Omg^{(0)}}{\Re T_{\rm rad}})
 (1-\frac{\lmd\Omg^{(L)}}{\Re T_{\rm rad}})}
 {(1+\frac{\lmd\Omg^{(0)}}{\Re T_{\rm rad}})
 (1+\frac{\lmd\Omg^{(L)}}{\Re T_{\rm rad}})}e^{-2\lmd}}, 
 \label{int:lncG}
\ee
we have 
\bea
 \Omg_{\rm eff}^{\rm 1loop} \eql 
 -\frac{9}{\pi^2}\int_0^\infty\dr x\;x\ln
 \brc{1-\frac{(1-6x\Omg^{(0)})(1-6x\Omg^{(L)})}{(1+6x\Omg^{(0)})(1+6x\Omg^{(L)})}
 e^{-12\Re T_{\rm rad} x}}+\cO(Q^2),  \label{Omg^1loop:simple}
\eea
where $x\equiv\frac{\lmd}{6\Re T_{\rm rad}}$. 
This agrees with (2.7) (or (6.32)) of Ref.~\cite{Rattazzi:2003rj} 
if we identify $-6\Omg^{(y_*)}$ in (\ref{Omg^1loop:simple}) 
with $\Omg_{y_*}$ in Ref.~\cite{Rattazzi:2003rj}. 

When $\Omg^{(0)}$ and $\Omg^{(L)}$ are small, 
$\Omg_{\rm eff}^{\rm 1loop}$ can be expanded as 
\be
 \Omg_{\rm eff}^{\rm 1loop} = \frac{\zeta(3)}{16\pi^2(\Re T_{\rm rad})^2}
 \brc{1-2\frac{\Omg^{(0)}+\Omg^{(L)}}{\Re T_{\rm rad}}
 +3\brkt{\frac{\Omg^{(0)}+\Omg^{(L)}}{\Re T_{\rm rad}}}^2}+\cdots. \label{ap:Omg^1loop}
\ee
We have used that 
\be 
 \int_0^\infty\dr\lmd\;\lmd^2\brkt{\coth\lmd-1} 
 = \frac{1}{3}\int_0^\infty\dr\lmd\;\frac{\lmd^3}{\sinh^2\lmd} 
 = 2\cZ(0) = \frac{\zeta(3)}{2}. \label{formula:zeta1}
\ee
Eq.(\ref{ap:Omg^1loop}) agrees with the expression, 
\be
 \Omg_{\rm eff}^{\rm 1loop} = \frac{\zeta(3)}{16\pi^2}
 \brkt{\Re T_{\rm rad}+\Omg^{(0)}+\Omg^{(L)}}^{-2}+\cdots, \label{ap:Omg1}
\ee
up to the quadratic order in $\Omg^{(y_*)}$. 

In the limit $\Omg^{(L)}\to\infty$, on the other hand, 
$\Omg_{\rm eff}^{\rm 1loop}$ is expanded as 
\be
 \Omg_{\rm eff}^{\rm 1loop} = -\frac{3}{64\pi^2(\Re T_{\rm rad})^2}
 \brc{1-2\frac{\Omg^{(0)}}{\Re T_{\rm rad}}
 +3\brkt{\frac{\Omg^{(0)}}{\Re T_{\rm rad}}}^2}+\cdots. 
 \label{ap2:Omg^1loop}
\ee
We have used that 
\be
 \int_0^\infty\dr\lmd\;\lmd^2\brkt{\tanh\lmd-1}
 = -\frac{1}{3}\int_0^\infty\dr\lmd\;\frac{\lmd^3}{\cosh^2\lmd}
 = -\frac{3}{8}\zeta(3). 
\ee
Eq.(\ref{ap2:Omg^1loop}) agrees with the expression, 
\be
 \Omg_{\rm eff}^{\rm 1loop} = -\frac{3\zeta(3)}{64\pi^2}
 \brkt{\Re T_{\rm rad}+\Omg^{(0)}}^{-2}+\cdots, \label{ap:Omg2}
\ee
up to the quadratic order in $\Omg^{(0)}$. 

The approximate expressions~(\ref{ap:Omg1}) and (\ref{ap:Omg2}) 
coincide with (5.5) and (5.6) in Ref.~\cite{Gregoire:2004nn}, respectively, 
if we assume that $\Omg^{(0)}=\Omg_0^{(0)}-\abs{q_0}^2/6$, 
where $\Omg_0^{(0)}$ is a real constant. 

\subsection{LARGE Volume Scenario in 5D SUGRA} 
In order to illustrate an impact of the non-geometric moduli 
on the moduli stabilization, 
we construct a simple model, in which the moduli 
are stabilized by $\Omg_{\rm eff}^{\rm 1loop}$ 
and an exponentially large extra dimension is dynamically realized.

\subsubsection{Approximate no-scale structure}
The moduli K\"ahler potential is rewritten as 
\be
 \Omg_{\rm eff} = -3\hat{\cN}^{1/3}+\frac{(n_V-\bar{n}_H+1)\zeta(3)}
 {32\pi^2\hat{\cN}^{2/3}}+\cdots, 
\ee
where the first and the second terms are the tree-level and the one-loop contributions 
respectively, the ellipsis denotes terms involving the matter fields, 
and the effective number of the hypermultiplets~$\bar{n}_H$ is defined as 
\be
 \bar{n}_H \equiv \sum_a n_a\frac{\cZ(d_a\cdot\Re T)}{\cZ(0)} 
 \leq n_H.  \label{def:barn_H}
\ee
This counts the number of hypermultiplets that spread over the bulk. 

Here we assume that $L_{\rm phys}=\vev{\hat{\cN}^{1/2}}\gg 1$. 
In this case, some moduli have very large VEVs, which are collectively denoted as $T_b$. 
The other moduli are denoted as $T_s$.  
Then the effective K\"ahler potential~$K$ is expanded as 
\be
 K = -3\ln\brkt{-\frac{\Omg_{\rm eff}}{3}} 
 = -\ln\hat{\cN}-\frac{\xi}{\hat{\cN}}+\cO\brkt{\frac{\xi^2}{\hat{\cN}^2}}, 
 \label{LVS:Kaehler}
\ee
where
\be
 \xi \equiv \frac{(\bar{n}_H-n_V-1)\zeta(3)}{32\pi^2}. 
\ee
This K\"ahler potential satisfies the following approximate no-scale relation, 
\be
 K_{\Io}K^{\Io\bar{J}_{\rm o}}K_{\bar{J}_{\rm o}} 
 = 3+\frac{6\xi}{\hat{\cN}}+\frac{4\xi_{\Io}\Re T^{\Io}}{\hat{\cN}}
 +\frac{\xi_{\Io}K^{\Io\bar{J}_{\rm o}}\xi_{\bar{J}_{\rm o}}}{\hat{\cN}^2}
 +\cO\brkt{\frac{\xi^2}{\hat{\cN}^2}},  \label{break:no-scale}
\ee
where $K^{\Io\bar{J}_{\rm o}}$ is an inverse matrix of the K\"ahler metric, and 
\be
 \xi_{\Io} \equiv \frac{\der\xi}{\der T^{\Io}} 
 = \sum_a\frac{n_a d_{a\Io}(d_a\cdot\Re T)}{16\pi^2}
 \ln\brkt{1-e^{-2\abs{d_a\cdot\Re T}}}. 
\ee
Note that $\abs{d_a\cdot\Re T}\gg 1$ unless $d_{aT_b}$ is negligibly small 
or a nontrivial cancellation occurs. 
Thus $\xi_{T_b}$ is exponentially small, and  
the $\xi$-dependent terms in (\ref{break:no-scale}) 
are all suppressed by $\hat{\cN}^{-1}$. 
Namely the no-scale structure is broken only by the corrections 
of $\cO(1/\hat{\cN})$. 


\subsubsection{Moduli stabilization}
Now we consider a specific model with two moduli 
multiplets~$(V_b,\Sgm_b)$ and $(V_s,\Sgm_s)$,  
and assume $\hat{\cN}$ as 
\be
 \hat{\cN}(\cV) = \cV_b^3-C_s\cV_s^3, 
\ee
where the constant~$C_s$ is assumed to be positive and typically $\cO(1)$.\footnote{
The 5D SUGRA description is not valid in the region of the moduli space 
in which $\vev{\hat{\cN}}\simeq 0$.  
} 
We further assume that the gaugino condensation occurs 
in a non-Abelian sector~$r=G$, in which $C^G_{T_b}=0$ and $C^G_{T_s}=\cO(1)$ 
in (\ref{eff:fW}). 
Then the following the effective superpotential~$W$ is induced.  
\be
 W(T) = W_0+A e^{-a T_s}, 
\ee
where the constants~$W_0$ and $A$ are of $\cO(1)$, and $a=\cO(4\pi^2)$. 
In this setup, we will show that there is a vacuum 
where $\Re\vev{T_b}\gg\Re\vev{T_s}$ and 
$L_{\rm phys}$ becomes exponentially large. 
Therefore, $T_b$ is almost identified as the radion 
and $T_s$ is the non-geometric modulus. 

The scalar potential is calculated as 
\bea
 V_{\rm pot} \eql e^K\brkt{D_I WK^{I\bar{J}}D_{\bar{J}}\bar{W}-3\abs{W}^2} \nonumber\\
 \eql \frac{1}{\hat{\cN}}\brc{\frac{2\hat{\cN}(aA)^2e^{-2a\tau_s}}
 {3C_s\tau_s\brkt{1-\frac{\dlt}{6}}}
 +4a\tau_s\brkt{1+\frac{\ep}{3}+\frac{\dlt}{6}}W_0A e^{-a\tau_s}\cos(a\rho_s)} \nonumber\\
 &&+\frac{6\xi W_0^2}{\hat{\cN}^2}\brc{1-\frac{2C_s\tau_s^3}{3\xi}
 \brkt{\ep-\frac{\dlt}{4}}}
 +\cdots,  \label{V_scalar}
\eea
where $D_IW\equiv W_I+K_IW$, $T_s\equiv\tau_s+i\rho_s$, 
the ellipsis denotes higher order in $\xi/\hat{\cN}$, and 
\be
 \ep \equiv \frac{\xi'}{C_s\tau_s^2}, \;\;\;\;\;
 \dlt \equiv \frac{\xi''}{C_s\tau_s}. 
\ee
Here the prime denotes the derivative with respective to $\tau_s$. 
By solving the minimization condition for $V_{\rm pot}$, 
we find the vacuum at the leading order in the $\xi/\hat{\cN}$-expansion as  
\bea
 \vev{\tau_b^3} \sma \vev{\hat{\cN}}  
 = \frac{3\xi W_0 e^{a\vev{\tau_s}}}{a\vev{\tau_s}A}
 \brc{1-\frac{\ep}{3}\brkt{1+\frac{2C_s\tau_s^3}{\xi}}
 -\frac{\dlt}{6}\brkt{1-\frac{C_s\tau_s^3}{\xi}}}, \nonumber\\
 \vev{\tau_s} \sma \brkt{\frac{\xi}{C_s}}^{1/3}
 \brc{1-\frac{2\ep}{3}\brkt{1+\frac{C_s\tau_s^3}{3\xi}}
 -\frac{\dlt}{6}\brkt{1-\frac{C_s\tau_s^3}{3\xi}}} \equiv \tau_s^{(0)}, \nonumber\\
 \cos\vev{\rho_s} \eql -{\rm sign}\brkt{W_0A},  \label{LVS:VEVs}
\eea
where $T_b\equiv\tau_b+i\rho_b$. 
Since $\rho_b$ does not appear in $V_{\rm pot}$, its VEV is not determined 
at this order. 
We have assumed that 
\be
 a\vev{\tau_s} \sim a\brkt{\frac{\xi}{C_s}}^{1/3} \gg 1,  \label{gg:atau_s}
\ee
which leads to an exponentially large $\hat{\cN}$.  
Hence the condition for the validity of the above analysis is (\ref{gg:atau_s}). 
When the standard model particles live in the bulk, 
the number~$n_H-n_V-1$ is around 40. 
Thus typical value of $\xi$ is of $\cO(0.1)$. 
In this case, (\ref{gg:atau_s}) is satisfied for $a=\cO(4\pi^2)$, 
and thus an exponentially large value of $\vev{\hat{\cN}}$ is obtained. 
For example, $L_{\rm phys}=\vev{\hat{\cN}^{1/2}}$
becomes $\cO(10^2)$, $\cO(10^5)$ and $\cO(10^{8})$ 
for $(\xi,C_s,W_0/A,a)=(0.1,1,1,4\pi^2)$, 
$(0.1,0.8,1,8\pi^2)$ and $(0.1,0.2,1,8\pi^2)$, respectively. 
Expanding the potential~(\ref{V_scalar}) around the vacuum~(\ref{LVS:VEVs}), 
we find that the moduli are non-tachyonic and have the following masses. 
\bea
 m_{\tau_b} \sma \Lvev{\frac{12\sqrt{6\xi}W_0}{\hat{\cN}\sqrt{a\tau_s}}
 \brc{1+\frac{C_s\tau_s^3}{9\xi}\brkt{\ep+\frac{11}{4}\dlt-\zeta}
 +\frac{1}{9}\brkt{-8\ep+3\dlt+\zeta}}}, \nonumber\\
 m_{\rho_b} \sma 0, \;\;\;\;\;
 m_{\tau_s},m_{\rho_s} \simeq \Lvev{\frac{4a\tau_sW_0}{\sqrt{\hat{\cN}}}
 \brc{1+\frac{\ep}{3}+\frac{\dlt}{12}}}, 
 \label{moduli_mass}
\eea
where $\zeta\equiv\xi'''/C_s$. 
Therefore (\ref{LVS:VEVs}) is a stable vacuum. 
It is a SUSY-breaking vacuum as we will see below. 
The potential value at this vacuum is 
\be
 V_{\rm min} \simeq -\Lvev{\frac{18\xi W_0^2}{(2a\tau_s+1)\hat{\cN}^2}}. 
\ee
Thus we need an extra source of SUSY-breaking 
to cancel this negative vacuum energy. 
Since $\abs{V_{\rm min}}$ is exponentially suppressed, 
the corrections to (\ref{LVS:VEVs}) and (\ref{moduli_mass}) 
by including such extra SUSY-breaking are negligible.

\subsubsection{Soft SUSY-breaking masses}
The gravitino mass is calculated as 
\be
 m_{3/2} = \vev{e^{K/2}W} \simeq \frac{W_0}{\sqrt{\vev{\hat{\cN}}}}. 
\ee
The F-term of $T_b$ is given by
\be
 \Lvev{\frac{F^{T_b}}{T_b+\bar{T}_b}} \simeq 
 -\Lvev{\frac{K^{T_b\bar{T}_b}D_{\bar{T}_b}\bar{W}
 +K^{T_b\bar{T}_s}D_{\bar{T}_s}\bar{W}}{2\tau_s\sqrt{\hat{\cN}}}} 
 \simeq \frac{W_0}{\sqrt{\vev{\hat{\cN}}}} \simeq m_{3/2}. 
\ee 
The F-term of $T_s$ vanishes at the leading order 
in the $\cO(1/(a\vev{\tau_s}))$-expansion. 
Thus we need to evaluate $\vev{\tau_s}$ including the next leading order. 
Then $\vev{\tau_s}$ in (\ref{LVS:VEVs}) is modified as 
\be
 \vev{\tau_s} = \tau_s^{(0)}+\frac{3}{2a^2\tau_s^{(0)}}
 \brc{1-\frac{\ep}{9}\brkt{2+\frac{5C_s\tau_s^{(0)3}}{\xi}}
 +\frac{\dlt}{9}\brkt{1+\frac{C_s\tau_s^{(0)3}}{\xi}}}. 
\ee
Here we used an assumption that $\vev{\ep},\vev{\dlt}\ll 1$, which is valid 
for typical values of the parameters. 
Then we obtain 
\be
 \Lvev{\frac{F^{T_s}}{T_s+\bar{T}_s}} \simeq 
 -\Lvev{\frac{K^{T_s\bar{T}_s}W_{\bar{T}_s}
 +K^{T_s\bar{I}}K_{\bar{I}}\bar{W}}{2\tau_s\sqrt{\hat{\cN}}}}
 \simeq \frac{m_{3/2}}{a\tau_s^{(0)}} 
 \ll \Lvev{\frac{F^{T_b}}{T_b+\bar{T}_b}}. 
\ee
The F-term of the compensator~$\phi_C$ is given by
\be
 \Lvev{\frac{F^{\phi_C}}{\phi_C}} = \frac{1}{3}\Lvev{K_IF^I}+m_{3/2} 
 \simeq \cO\brkt{\frac{m_{3/2}}{\hat{\cN}^{4/3}}}  
 \ll \Lvev{\frac{F^{T_s}}{T_s+\bar{T}_s}},  
\ee
where the cancellation of the leading contributions is ensured by 
the approximate no-scale structure of the K\"ahler 
potential~\cite{ArkaniHamed:2004yi}. 
The F-terms of the other chiral superfields are negligible. 
Therefore, the dominant source of SUSY breaking is the F-term of $T_b$. 

Now let us consider the soft SUSY-breaking masses. 
Since the gauge coupling constants~$g_r$ $(r=U(1)_Y,SU(2)_L,SU(3)_c,\cdots)$ 
are determined by $\frac{1}{g_r^2}=\Re f_{\rm eff}^r$, 
where $f_{\rm eff}^r$ is given in (\ref{eff:fW}), 
$C_{T_b}^r$ must be zero or negligible. 
Otherwise $g_r$ become much smaller than the observed values.  
Thus the gaugino masses~$M_r$  are evaluated as 
\be
 M_r = \Lvev{F^I\der_I\ln\brkt{\Re f_{\rm eff}^r}} 
 \simeq \Lvev{\frac{F^{T_s}}{T_s+\bar{T}_s}} 
 \simeq \frac{m_{3/2}}{a\tau_s^{(0)}}.  
\ee

As for the matter multiplets, there are two possibilities, 
\ie, they are in the bulk or localized on the boundaries. 
The soft scalar masses in each case are estimated as follows. 
\begin{description}
\item[Bulk matter] \mbox{} \\
Since the K\"ahler potential for the bulk matters are read off from (\ref{multi:flat}) as 
\be
 \Omg_{\rm matter} = \sum_a 2\hat{\cN}^{1/3}Y_{d_a}\abs{Q_a}^2+\cdots, 
\ee
the SUSY-breaking scalar masses of $Q_a$ are calculated as
\bea
 m_{Q_a}^2 \eql -\Lvev{F^I\bar{F}^{\bar{J}}\der_I\der_{\bar{J}}
 \ln\brkt{2\hat{\cN}^{1/3}Y_{d_a}}} \nonumber\\
 \sma m_{3/2}^2\brc{1-(d_a\cdot\Re\vev{T})^2\cY(d_a\cdot\Re\vev{T})}, 
\eea
where
\be
 \cY(x) \equiv \frac{1+e^{4x}-2e^{2x}(1+2x^2)}{(1-e^{2x})^2x^2},  
\ee
is monotonically decreasing function of $\abs{x}$ and $\cY(0)=1/3$. 
Since $\lim_{x\to\infty}x^2\cY(x)=1$, 
these masses become much smaller than $m_{3/2}$ 
when the wave function for $Q_a$ is strongly localized toward one of the boundaries, 
\ie, $\abs{d_a\cdot\Re\vev{T}}\gg 1$.

\item[Brane matter] \mbox{} \\
The brane matter does not couple with the moduli at tree level. 
Thus we need to take into account the one-loop contributions 
in order to estimate the soft SUSY-breaking masses for them. 
From (\ref{Omg^1loop}) with (\ref{def:cGs}) and (\ref{def:cHs}), we obtain 
\be
 \Omg_{\rm eff} = \brkt{\Omg^{(0)}+\Omg^{(L)}}
 -\frac{\zeta(3)}{8\pi^2\hat{\cN}}\brkt{\Omg^{(0)}+\Omg^{(L)}}+\cdots, 
\ee
where the ellipsis denotes terms independent of the brane-localized fields 
or higher order terms in the $\hat{\cN}^{-1}$-expansion. 
The first term is the tree-level contribution. 
We have used (\ref{formula:zeta1}) to obtain the second term. 
As an example, we consider a case that
\be
 \Omg^{(0)} = \Omg_0^{(0)}+h_q\abs{q_0}^2, \;\;\;\;\;
 \Omg^{(L)} = 0, 
\ee
where $\Omg_0^{(0)}$ and $h_q$ are constants. 
Then the soft mass for $q_0$ is computed as 
\be
 m_q^2 = -\Lvev{F^I\bar{F}^{\bar{J}}\der_I\der_{\bar{J}}
 \ln\brc{h_q\brkt{1-\frac{\zeta(3)}{8\pi^2\hat{\cN}}}}} 
 \simeq \frac{3\zeta(3)}{2\pi^2\vev{\hat{\cN}}}m_{3/2}^2, 
\ee
which is much smaller than the soft masses for the bulk matters.
\end{description}

The mass scales of this model in the unit of $M_{\rm Pl}$ are 
summarized in Table~\ref{Mass_spectrum}. 
We have assumed that $W_0=\cO(1)$ there. 
\begin{table}[t]
\begin{center}
\begin{tabular}{c|c|c|c|c} \hline
$M_{\rm Pl}$ & $m_{\tau_s}$ & $m_{\rho_s}$ & $m_{\rm KK}$ & $m_{3/2}$ 
\\ \hline
1 & $\cO\brkt{\frac{\ln L_{\rm phys}}{L_{\rm phys}}}$ & 
$\cO\brkt{\frac{\ln L_{\rm phys}}{L_{\rm phys}}}$ 
& $\cO\brkt{\frac{1}{L_{\rm phys}}}$ 
& $\cO\brkt{\frac{1}{L_{\rm phys}}}$ \\ \hline\hline
$m_{Q_a}$ & $M_r$ & $m_{\tau_b}$ & $m_q$ & $m_{\rho_b}$ \\ \hline 
$\simlt\cO\brkt{\frac{1}{L_{\rm phys}}}$ & 
$\cO\brkt{\frac{1}{L_{\rm phys}\ln L_{\rm phys}}}$ & 
$\cO\brkt{\frac{1}{L_{\rm phys}^2}}$ & 
$\cO\brkt{\frac{1}{L_{\rm phys}^2}}$ & $\simeq 0$  
\\ \hline
\end{tabular}
\end{center}
\caption{The orders of magnitude of the mass eigenvalues 
in the unit of $M_{\rm Pl}$. 
The size of the extra dimension~$L_{\rm phys}$ is exponentially large in our model. 
}
\label{Mass_spectrum}
\end{table}

Note that the above spectrum is basically that of 
the Scherk-Schwarz SUSY-breaking~\cite{Scherk:1978ta} 
because the dominant SUSY-breaking source is provided by the F term of 
the radion superfield~$T_{\rm rad}\simeq T_b$~\cite{Abe:2005wn,Gersdorff}. 
In our model, an exponentially large extra dimension is dynamically realized 
with the aid of the non-geometric moduli~$T_s$.

\subsubsection{Comparison with LARGE volume scenario in string theory}
Finally let us compare the LARGE volume scenario 
in type IIB string theory~\cite{Balasubramanian:2005zx,Conlon:2005ki}. 
In this scenario, the K\"ahler potential for the K\"ahler moduli has a structure, 
\be
 K = -2\ln\brkt{V_{\rm CY}+\frac{\xi}{2}}+\cdots, \label{K:string}
\ee
where $V_{\rm CY}$ is the volume of 6-dimensional compact space~$M$ 
in the string frame, 
$\xi=-\frac{\chi(M)\zeta(3)}{2(2\pi)^3}=0.48$ ($\chi$ is the Euler number).  
The ellipsis denotes terms dependent on the other moduli. 
On the other hand, the moduli K\"ahler potential in our model~(\ref{LVS:Kaehler}) 
can be rewritten as 
\be
 K = -3\ln\brkt{\hat{\cN}^{1/3}+\frac{\xi}{3\hat{\cN}^{2/3}}}
 +\cO\brkt{\frac{\xi^2}{\hat{\cN}^2}}.  \label{K:5DSUGRA}
\ee
In both (\ref{K:string}) and (\ref{K:5DSUGRA}), the K\"ahler potential has the no-scale structure at the leading order,\footnote{
For the explicit moduli-dependence of $V_{\rm CY}$, 
see (6) and (7) in Ref.~\cite{Conlon:2005ki}. }  
and the subleading term proportional to $\xi$ breaks it. 
However, the origin of $\xi$ is different in the two cases. 
In (\ref{K:string}), it comes from the $\alp'$-correction, that is a stringy effect. 
In (\ref{K:5DSUGRA}), it is induced by the one-loop correction and 
thus obtained within the field theory. 
Besides, $\xi$ in (\ref{K:string}) is a constant 
while it depends on the moduli in (\ref{K:5DSUGRA}). 

The mass spectrum in Ref.~\cite{Balasubramanian:2005zx,Conlon:2005ki} is 
\bea
 m_{\rm KK} \eql \cO\brkt{\frac{1}{\vev{V_{\rm CY}^{2/3}}}}, \;\;\;\;\;
 m_{\tau_s},m_{\rho_s} = \cO\brkt{\frac{\ln\vev{V_{\rm CY}}}{\vev{V_{\rm CY}}}}, 
 \nonumber\\
 m_{3/2},m_S,m_\phi \eql \cO\brkt{\frac{1}{\vev{V_{\rm CY}}}}, \;\;\;\;\;
 m_{\tau_b} = \cO\brkt{\frac{1}{\vev{V_{\rm CY}^{3/2}}}}, \;\;\;\;\;
 m_{\rho_b} \simeq 0.  \label{spectrum:LVS}
\eea
The moduli~$\tau_b$ and $\tau_s$ correspond to a large and a small cycles  
in a `Swiss-cheese' structure of the Calabi-Yau manifold. 
The moduli~$\rho_b$ and $\rho_s$ are their axionic partners.   
The other moduli~$S$ and $\phi$ are the dilaton-axion and the complex structure moduli 
respectively.  
Comparing our spectrum in Table~\ref{Mass_spectrum} with (\ref{spectrum:LVS}), 
we find that the nongeometric moduli~$\tau_s$ and $\rho_s$ are heavier 
than the KK mass scale~$m_{\rm KK}=\pi/\vev{\hat{\cN}^{1/2}}$ 
in contrast to (\ref{spectrum:LVS}). 
Thus the expressions of their masses in our analysis are valid 
only when $W_0<\cO(1/a\vev{\tau_s})$, 
although our mechanism that realizes a large extra dimension still works 
even if $W_0=\cO(1)$.

\section{Summary} \label{summary}
We discussed the impacts of the non-geometric moduli 
on 4D effective theory of 5D SUGRA on $S^1/Z_2$. 
Such moduli often exist when we construct models based on generic 5D SUGRA. 

At tree level, additional matter quartic terms are induced  
in the effective K\"ahler potential 
by integrating out the non-geometric moduli, 
and they can significantly affect the flavor structure of the sfermions 
for the bulk matters, as we pointed out in Ref.~\cite{Abe:2008an}. 
In the flat spacetime, the moduli K\"ahler potential 
has the no-scale structure, and thus the potential for the moduli is not 
generated. 
The warped geometries are obtained by gauging an isometry 
on the hyperscalar manifold with the moduli multiplets~${\mathbb V}^{\Io}$. 
This corresponds to a case that the compensator multiplet is charged 
for ${\mathbb V}^{\Io}$ in our off-shell formulation. 
Notice that such warped geometries generically deviate from the familiar 
Randall-Sundrum spacetime in the multi moduli case 
because the VEVs of the non-geometric moduli also 
contribute to the geometry. 
The Randall-Sundrum geometry is just a special limit 
in the multi moduli case, 
which is realized when the isometry is gauged only by the graviphoton 
(or radion) multiplet. 

At one-loop level, the no-scale structure in the flat spacetime is broken.  
Thus the moduli have a nontrivial potential, and can be stabilized. 
This is interpreted as the stabilization 
by the Casimir effect~\cite{Fabinger:2000jd,Garriga:2000jb,Toms:2000bh}. 
The one-loop K\"ahler potential in the multi moduli case is calculated 
in our previous work~\cite{Sakamura:2013wia}, 
including generic form of the norm function and the boundary-localized terms. 
We checked that this result is consistent with those of other related 
works~\cite{Falkowski:2005fm,Rattazzi:2003rj,Gregoire:2004nn}, 
which were obtained in a simple case, \ie, 
ungauged SUGRA without the non-geometric moduli. 

To illustrate the impact of the non-geometric moduli, 
we also construct a simple model, in which the size of the extra dimension 
is stabilized at an exponentially larger value than the Planck length. 
This dynamical realization of the large extra dimension 
is similar to the LARGE volume scenario in string theory. 
In contrast to the latter, we should note that 
the correction to the no-scale structure, 
which is a key of this scenario, 
is obtained within the field theory. 
Since the subleading corrections are suppressed by inverse powers of 
the large extra dimension, the results obtained here is robust. 
This scenario works thanks to the existence of the non-geometric moduli. 
The dominant source of SUSY breaking is provided by the F term of the radion 
superfield so the spectrum is essentially that of the Scherk-Schwarz SUSY breaking. 
Detailed phenomenological analysis of this model is interesting, 
and we will leave it for a future publication. 

In this paper, we focused on the case that the gauge groups for the moduli 
multiplets are Abelian, for simplicity. 
When the moduli and gauge supermultiplets, ${\mathbb V}^{\Io}$ 
and ${\mathbb V}^{\Ie}$, form a non-Abelian gauge multiplet, 
the spontaneous breaking of the gauge symmetry can occur 
by the Hosotani mechanism~\cite{Hosotani:1983xw} 
and the moduli will form gauge multiplets under the unbroken gauge group. 
Thus we can discuss, for example, the gauge-Higgs unification scenario 
at the grand unification scale~\cite{Hebecker:2003jt,Haba:2002vc} 
after extending our formula~(\ref{Omg^1loop}) to the non-Abelian case.  
This issue is also left for a future work.

\subsection*{Acknowledgements}
The authors would like to thank Hiroyuki Abe and 
Tetsutaro Higaki for useful information and discussions. 
This work was supported in part by 
Grant-in-Aid for Scientific Research (C) No.25400283  
from Japan Society for the Promotion of Science (Y.S.),   
and a Grant for Excellent Graduate Schools, MEXT, Japan (Y.Y.). 

\appendix

\section{Matrices constructed from the norm function} \label{matrices}
The vector sector is characterized by the norm function~$\cN(X)$ 
defined in (\ref{def:norm_fcn}). 
The coefficients of the kinetic terms for the vector multiplets are 
given by~\cite{Kugo:2000af} 
\be
 a_{IJ} \equiv -\frac{1}{2\cN}\brkt{\cN_{IJ}-\frac{\cN_I\cN_J}{\cN}}, 
 \label{def:a_IJ}
\ee
where $\cN_I\equiv\der\cN/\der X^I$ 
and $\cN_{IJ}\equiv\der^2\cN/\der X^I\der X^J$. 
This matrix is positive definite for physically sensible theories. 

One combination of the vector multiplets is identified with 
the graviphoton superfield, 
\be
 V_G \equiv \Lvev{\frac{\cN_I}{3\cN^{2/3}}}V^I.  \label{def:V_G}
\ee
This is essentially auxiliary degree of freedom in our superfield formalism. 
Thus we define the following projection operator~$\cP_V$ 
that eliminate $V_G$ from the $n_V$ vector superfields~\cite{Kugo:2000af}. 
\be
 (\cP_V)^I_{\;\;J}(X) \equiv \dlt^I_{\;\;J}-\frac{X^I\cN_J}{3\cN}. 
 \label{def:cP_V}
\ee
This satisfies 
\be
 \cN_I(\cP_V)^I_{\;\;J} = (\cP_V)^I_{\;\;J}X^J = 0, \;\;\;\;\;
 \cP_V^2 = \cP_V. 
\ee

For the norm function~(\ref{mono_norm}), for example, the matrices~(\ref{def:a_IJ}) 
and (\ref{def:cP_V}) become 
\be
 a(X) = \begin{pmatrix} \frac{1}{(X^1)^2} & 0 \\ 0 & \frac{1}{2(X^2)^2} 
 \end{pmatrix}, \;\;\;\;\;
 \cP_V(X) = \frac{1}{3}\begin{pmatrix} 1 & -\frac{X^1}{X^2} \\
 -\frac{2X^2}{X^1} & 2 \end{pmatrix}.  \label{expr:acP_V}
\ee
Thus, 
\be
 \cP_V a^{-1} = \frac{1}{3}\begin{pmatrix} (X^1)^2 & -2X^1X^2 \\
 -2X^1X^2 & 4(X^2)^2 \end{pmatrix}. 
\ee

\section{Functions for $\bdm{\Omg_{\rm eff}^{\rm 1loop}}$} \label{def:functions}
The functions in (\ref{Omg^1loop}) are defined in terms of the quantities in 
$\cL^{(y_*)}_{\rm bd}$ in (\ref{cL_bd}) as 
\ignore{
\bea
 \cG_U(\lmd) \eql 1+\frac{\lmd^2\Omg^{(0)}\Omg^{(L)}}{\hat{\cN}^{2/3}}
 +\frac{\lmd}{\hat{\cN}^{1/3}}\brkt{\Omg^{(0)}+\Omg^{(L)}}\coth\lmd, \nonumber\\
 \cG_V(\lmd) \eql \det\brc{\id_{n_{V_{\rm e}}}
 +\frac{\lmd^2}{\hat{\cN}^{2/3}}H_V^{(0)}H_V^{(L)}
 +\frac{\lmd}{\hat{\cN}^{1/3}}\brkt{H_V^{(0)}+H_V^{(L)}}\coth\lmd}, \nonumber\\
 \cG_{\rm ch}(\lmd) \eql \det\left\{2e^{-\frac{T_R}{2}-\lmd}
 \right.
\eea
}
\bea
 \cG_U(\lmd) \eql 1+\frac{\cN^{2/3}}{\lmd^2\Omg^{(L)}\Omg^{(0)}}
 +\frac{\cN^{1/3}}{\lmd}\brkt{\frac{1}{\Omg^{(L)}}+\frac{1}{\Omg^{(0)}}}\coth\lmd, 
 \nonumber\\
 \cG_V(\lmd) \eql \det\left\{\id_{n_{V_e}}
 +\frac{\cN^{2/3}}{\lmd^2}H_V^{(L)-1}H_V^{(0)-1}
 +\frac{\cN^{1/3}}{\lmd}\brkt{H_V^{(L)-1}+H_V^{(0)-1}}\coth\lmd\right\}, 
 \nonumber\\
 \cG_{\rm ch}(\lmd) \eql \det\left\{2e^{-\frac{T_R}{2}-\lmd}\sinh\omg_T
 +\frac{2\cN^{2/3}}{\lmd^2}H_{\rm ch}^{(L)-1}
 e^{\frac{T_R}{2}-\lmd}\sinh\omg_T H_{\rm ch}^{(0)-1} \right. \nonumber\\
 &&\hspace{10mm}
 +\frac{2\cN^{1/3}}{\lmd^2}H_{\rm ch}^{(L)-1}e^{\frac{T_R}{2}-\lmd}
 \brkt{\omg_T\cosh\omg_T-\frac{T_R}{2}\sinh\omg_T} \nonumber\\
 &&\hspace{10mm}\left.
 +\frac{2\cN^{1/3}}{\lmd^2}e^{-\frac{T_R}{2}-\lmd}
 \brkt{\omg_T\cosh\omg_T+\frac{T_R}{2}\sinh\omg_T}H_{\rm ch}^{(0)-1}\right\}
 \nonumber\\
 &&\times\brc{\det\brkt{2e^{-\frac{T_R}{2}-\lmd}\sinh\omg_T}}^{-1},  \label{def:cGs}
\eea
and 
\bea
 \cH_U^{(y_*)}(\lmd) \defa 1+\frac{\cN^{1/3}}{\lmd \Omg^{(y_*)}}, \;\;\;\;\;
 \cH_V^{(y_*)}(\lmd) \equiv \det\brkt{\id+\frac{\cN^{1/3}}{\lmd}H_V^{(y_*)-1}}, 
 \nonumber\\
 \cH_{\rm ch}^{(0)}(\lmd) \defa \det\brc{\id+\frac{\cN^{1/3}}{\lmd^2}
 \brkt{\omg_T+\frac{T_R}{2}}H_{\rm ch}^{(0)-1}}, \nonumber\\
 \cH_{\rm ch}^{(L)}(\lmd) \defa \det\brc{\id+\frac{\cN^{1/3}}{\lmd^2}
 H_{\rm ch}^{(L)-1}e^{T_R}\brkt{\omg_T-\frac{T_R}{2}}}.  \label{def:cHs}
\eea
Here the matrices~$H_V^{(y_*)}$, $H_{\rm ch}^{(y_*)}$, $T_R$ and $\omg_T$ 
are defined as 
\bea
 &&\brkt{H_V^{(y_*)}}^{\Ie}_{\;\;\Je} \equiv 
 a^{\Ie\Ke}\brkt{\frac{\Re f_{\Ke\Je}^{(y_*)}}{\cN^{2/3}}
 -\frac{2\Omg^{(y_*)}_{\Ke\Je}}{3\lmd^2}}, \nonumber\\
 &&\brkt{H_{\rm ch}^{(y_*)}}_{AB} \equiv \frac{1}{2}\brkt{\Omg_{\bar{A}B}^{(y_*)}
 +\frac{i\cN^{1/3}}{\lmd}W_{AB}^{(y_*)}}, \nonumber\\
 &&T_R \equiv \bigoplus_a\brkt{-2d_a\cdot\Re T\otimes\id_{n_a}}, \;\;\;\;\;
 \omg_T \equiv \brkt{\lmd^2+\frac{T_R^2}{4}}^{1/2},  
\eea
where indices~$A,B=1,\cdots,n_H$ run over all the physical hypermultiplets, 
and the suffixes of $\Omg^{(y_*)}$, $f^{(y_*)}$ and $W^{(y_*)}$ denote 
the derivatives with respect to the corresponding superfields.


\end{document}